# Applying the CobiT® Control Framework to Spreadsheet Developments

Raymond J Butler, CISA

H M Customs & Excise Computer Audit Service National Office, Queens Dock, Liverpool L74 4AA UK

✆ ++44 (0)151 703 8741 ray.butler@hmce.gsi.gov.uk

**ABSTRACT**

*One of the problems reported by researchers and auditors in the field of spreadsheet risks is that of getting and keeping management's attention to the problem. Since 1996, the Information Systems Audit & Control Foundation and the IT Governance Institute have published CobiT® which brings mainstream IT control issues into the corporate governance arena. This paper illustrates how spreadsheet risk and control issues can be mapped onto the CobiT framework and thus brought to managers' attention in a familiar format.*

## 1. A BRIEF INTRODUCTION TO COBIT®

### 1.1. What is CobiT ?

CobiT®, Control Objectives for Information & related Technology is a tool set which helps business managers to understand and manage the risks associated with implementing new technologies, and demonstrate to regulators, shareholders and other stakeholders how, and how well they have done this. It is based on international best practice in IT management and control.

The tool set facilitates IT governance, defined as "a structure of relationships and processes to direct and control the enterprise in order to achieve the enterprise's goals by adding value while balancing risk versus return over IT and its processes" [ISACF 2000(1)] In an age where business is almost entirely dependent on technology, IT Governance is an essential element of wider corporate governance.

### 1.2. CobiT's Contents

The framework defines

- 34 IT processes in 4 broad groups. These processes depend on and impact on IT resources.

- High-level control objectives for each of the 34 processes,

- 318 detailed control objectives, and associated audit guidelines.

CobiT also contains Management Guidelines, including Maturity Models, Critical Success Factors, Key Goal Indicators and Key Performance Indicators for each of the 34 processes.

### 1.3. CobiT's Audience: Management, Users And Auditors

The framework is designed to help three distinct audiences:

---

- **Management** – who need to balance risk and control investment in an IT environment which is often unpredictable.

- **Users** – who need to obtain assurance on the security and controls of the IT services they depend on to deliver their products and services to internal and external customers.

- **Auditors** – who can use it to substantiate their opinions and / or provide advice to management on internal controls.

Apart from responding to the needs of the immediate audience of senior management, auditors and security and control professionals, CobiT can be used within enterprises by business process owners in meeting their responsibility for control over the information aspects of their processes and by those responsible for IT in the enterprise.

## 2. HOW DOES COBIT COVER SPREADSHEET RISKS ?

No *specific* mention is made of Spreadsheets, or of end-user computing. Instead, CobiT provides a generic framework for *all* the principal IT processes. These can be adapted, scaled and applied to IT solutions at all levels, from a whole Enterprise Resource Planning system to a (relatively) simple spreadsheet development. The example below shows how this can be done.

### 2.1. An Example from CobiT

The high-level control objective for the process defined as "Acquire and Maintain Application Software" states [ISACF 2000(2)] that

**"Control over the IT process** of acquiring and maintaining application software **that satisfies the business requirement** to provide automated functions which effectively support the business process **is enabled by** the definition of specific statements of functional and operational requirements, and a phased implementation with clear deliverables, **and takes into consider**ation

- functional testing and acceptance
- application controls and security requirements
- documentation requirements
- application software life cycle
- enterprise information architecture
- system development life cycle methodology
- user-machine interface
- package customisation"

This is supported by 17 detailed control objectives covering

- Design Methods
- Major Changes to Existing Systems
- Design Approval
- File Requirements Definition and Documentation
- Programme Specifications
- Source Data Collection Design
- Input Requirements Definition and Documentation
- Definition of Interfaces
- User-Machine Interface
- Processing Requirements Definition and Documentation
- Output Requirements Definition and Documentation
- Controllability
- Availability as a Key Design Factor
- IT Integrity Provisions in Application Programme Software

---

- Application Software Testing
- User Reference and Support Materials
- Reassessment of System Design

All of these controls can be scaled and applied to spreadsheet development. In our "ideal environment", some of these (such as design approval, and testing) will require specific formal controls and procedures, some (such as availability) may depend on wider Office Technology platform controls

## 3. THE MATURITY MODEL

CobiT's maturity model for control over IT processes provides a method of scoring which enables an organisation to grade its IT control procedures on a scale from 0 (non-existent) to 5 (optimised). This approach has been derived from the Maturity Model for software development capability defined by the Software Engineering Institute. Management use the maturity model to map the current status of:

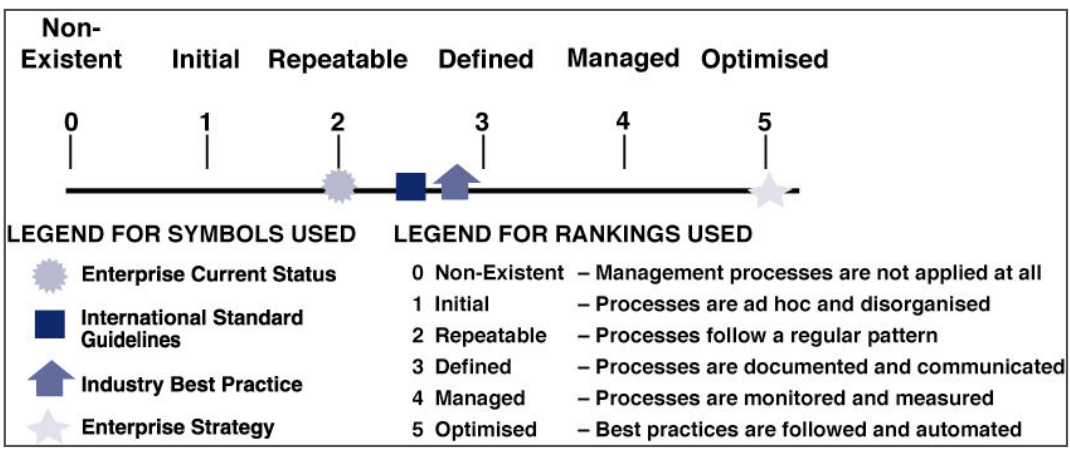

- their organisation,
- the best practice or the general state of practice in their industry
- international standards

**Figure 1 - The CoBIT ® Maturity Model**

and define where the organisation wants to be against these levels. Figure 1 [ISACF 2000(2)] illustrates the rankings and the way in which an organisation can use the model to map the maturity of their current and desired practices onto the model.

## 4. APPLYING CobiT TO SPREADSHEETS

The framework can be readily scaled to spreadsheet developments. The following control objectives (high level and detailed) and maturity model is offered as

- A demonstration of the adaptability of CobiT to spreadsheets, and
- a "first draft" upon which a formal set of overall and more detailed control objectives can be built.

The controls will obviously need to be applied only to the degree justified by the actual or potential impact that a spreadsheet model has upon the organisation in which it is used. A simple impact assessment and the contents of documentation, etc. have been described in previous papers by this author [Butler, 2000] and will not be reproduced here.

### 4.1. Control Objective

**Control** over the process of **developing and maintaining spreadsheet models and applications** that **satisfy the business requirement** to provide accurate and error-free business models and analyses which **effectively support the business process** is enabled by the definition of specific statements of functional and operational requirements, and a phased implementation with clear deliverables, and **takes into consideration**

Design Methods

Security and data retention requirements

Testing and Acceptance

Documentation Requirements

### 4.2. Detail Control Objectives :

**Design Methods**

The organisation should employ a spreadsheet development methodology which requires that appropriate procedures and techniques, involving close liaison with model users, are applied to create the design specifications for each new spreadsheet development and to verify the design specifications against the user requirements.

**Major Changes to Existing Systems**

Management should ensure, that in the event of major changes to existing spreadsheet models or applications, a similar development process is observed as in the case of the development of new models.

**Design Approval**

The organisation's spreadsheet development methodology should require that the design specifications for all spreadsheet development and modification projects be reviewed and approved by management, the affected user departments and the organisation's senior management, when appropriate.

**Programme Specifications**

The organisation's spreadsheet development methodology should require that detailed written specifications be prepared for each spreadsheet development or modification project. The methodology should further ensure that specifications agree with design specifications.

**Testing**

Testing to ensure that :

- The spreadsheet calculations

- Data input and controls over data

- Links between host systems and the spreadsheet, between parts of the spreadsheet and between spreadsheets in a multi-file suite of models

- Output reports

operate correctly and as specified according to the development test plan and established testing standards should be performed and documented before the development is approved by the user. Adequate measures should be conducted to prevent disclosure of sensitive information used during testing.

**User Documentation and instructions**

The organisation's spreadsheet development methodology should provide that adequate user reference and support manuals be prepared (preferably in electronic format) as part of every spreadsheet development or modification project.

**Security and retention**

- The organisation's spreadsheet development and use methodology should include directions for ensuring that :

- Access to spreadsheet models is restricted to authorised persons;

- Spreadsheet models are protected against inadvertent or unauthorised modification

- Spreadsheet models and applications are retained in electronic form for the period of time appropriate to the purpose of the spreadsheet.

### 4.3. Maturity Model

Control over the process of developing and maintaining spreadsheet models and applications that satisfy the business requirement to provide accurate and error-free business models and analyses which effectively support the business process

| Maturity Level | Characteristics |
| --- | --- |
| 0     Non-existent | There is no process for designing and specifying spreadsheets. Typically, spreadsheets are developed in an unstructured manner by untrained end-users, with little or no documentation of actual requirements and no testing. There is an extremely high risk of error in important spreadsheets. |
| 1     Initial/Ad Hoc | There is an awareness that a process for developing spreadsheets is required. Approaches, however, vary from development to development without any consistency and typically in isolation from each other. The organisation's business depends upon a variety of individual solutions with varying degrees of documentation and control and now suffers legacy problems and inefficiencies with maintenance and support. There is a very high risk of errors in important spreadsheets. |
| 2     Repeatable but Intuitive | There are similar processes for developing and maintaining spreadsheets, but they are based on the expertise within the users, not on a documented process. The success rate with spreadsheets depends greatly on individual users' skills and experience levels. Maintenance is usually problematic and suffers when internal knowledge has been lost from the organisation. There is a high risk of errors in important spreadsheets |
| 3     Defined Process | There are documented development and maintenance processes. An attempt is made to apply the documented processes consistently across different spreadsheet developments, but they are not always found to be practical to implement. They are generally inflexible and hard to apply in all cases, so steps are frequently bypassed. As a consequence, spreadsheets are often developed and implemented in a piecemeal fashion. Maintenance follows a defined approach, but is often time-consuming and inefficient. There is medium risk of errors in important spreadsheets. |
| 4     Managed and Measurable | There is a formal, clear and well-understood spreadsheet development and implementation methodology and policy that includes a formal design and specification process, a process for testing and requirements for documentation, ensuring that all spreadsheets are developed and maintained in a consistent manner. Formal approval mechanisms exist to |

| Maturity Level | Characteristics |
|---|---|
| | ensure that all steps are followed and exceptions are authorised. The methods have evolved so that they are well suited to the organisation and are likely to be positively used by all staff, and applicable to most important spreadsheet developments. There is a low risk of errors in important spreadsheets. |
| 5      Optimised | Spreadsheet development and maintenance practices are in line with the agreed processes. The development and maintenance process is well advanced, enables rapid deployment and allows for high responsiveness, as well as flexibility, in responding to changing business requirements. The spreadsheet development and implementation process has been subjected to continuous improvement and is supported by internal and external knowledge databases containing reference materials and best practices. The methodology creates computer based documentation in a pre-defined structure that makes production and maintenance very efficient. There is a very low risk of errors in important spreadsheets |

## 5. CONCLUSIONS

### 5.1. Can the CobiT Approach Help with Spreadsheets ?

As illustrated above, the CobiT approach can easily be applied to spreadsheets. In itself, this adds no new insights into the problem of spreadsheet risk, or into good practice and control issues. It will, however be a very useful method of presenting those issues. It will allow spreadsheet risk, good practice and control to be presented to managers in a familiar format. It will therefore help the audit and spreadsheet development communities to "market" the issues to decision makers, and raise them as corporate and IS governance rather than parochial technical issues.

### 5.2. What Next ?

Business managers and auditors urgently need two (linked) products. These are

- a brief synopsis of spreadsheet risks, to explain why they should take spreadsheets in their organisation seriously.

- A statement of good practice in the design, use and control of spreadsheets.

It is hoped that the proceedings of this and the previous EuSpRIG conference will provide much of the source material for this, and that this paper will influence its production on the CobiT format, increasingly familiar to and used by business managers, our intended audience.

## 6. RESOURCES AND REFERENCES

Much of CobiT is available as an open standard for download from the Information Systems Audit and Control Association web site at www.isaca.org

CobiT 3rd Edition Executive Summary, IS Audit & Control Foundation, Chicago, Ill, July 2000 [ISACF 2000(1)]

CobiT 3rd Edition Control Objectives, IS Audit & Control Foundation, Chicago, Ill, July 2000 [ISACF 2000(2)]

CobiT 3rd Edition Management Guidelines, IS Audit & Control Foundation, Chicago, Ill, July 2000 [ISACF 2000(3)]

Butler, R (2000) Is this Spreadsheet a Tax Evader ?", proceedings of the 33$^{rd}$ Hawaii International Conference on System Sciences

Butler, R (2000) "Risk Assessment in Spreadsheet Developments", Proceedings of the first EuSpRIG Conference